\documentclass[fleqn,twoside]{article}

\usepackage{espcrc2}

\usepackage{amssymb}

\usepackage{graphicx}
\usepackage{epsfig}

\usepackage[figuresright]{rotating}

\title{The Galactic Center as a point source of neutrons at EeV energies}

\author{G. Medina-Tanco\address[]{Instituto de Ciencias Nucleares, UNAM, Circuito Exterior
S/N, Ciudad Universitaria, Mexico D.F. 04510, Mexico}%
        \thanks{gustavo.medinatanco@nucleares.unam.mx},
        A. G. O. Krone-Martins\address{Instituto de Astronomia, Geof\'isica e Ci\^encias
Atmosf\'ericas, USP, Rua do Matao 1226, CEP 05508-900, Brasil} }

\begin{document}

\begin{abstract}
The central region of our Galaxy is a very peculiar environment,
containing magnetic fields in excess of 100 mG and gas densities
reaching $\sim10^4$cm$^{-3}$. This region was observed as a strong
source of GeV and TeVs gammas, what suggests that a mechanism of
proton-neutron conversion could be taking place therein. We
propose that the Galactic Center must also be a source of EeV
neutrons due to the conversion of ultra high energy cosmic ray
protons into neutrons via p-p interactions inside this region.
This scenario should be falsifiable by the Pierre Auger
Observatory after a few years of full exposure. \vspace{1pc}
\end{abstract}

\maketitle

\section{Introduction}

Cosmic rays at the highest energies, i.e., above few $\times
10^{17}$ eV, are very likely extragalactic and pose some of the
most intriguing questions of the contemporaneous research in
astrophysics. Their origin, composition, acceleration and
production mechanisms are still unclear \cite{nw00}. The first
step towards answering these questions will very likely require
the unambiguous identification of an astrophysical counterpart to
a subset (cluster) of events.

Despite the fact that, over wide angles on the sky, all
experiments agree that the flux of ultra-high energy cosmic rays
(UHECR) seems isotropic, there have been claims in the past for
the detection of anisotropy on smaller scales. The unambiguous
determination of the existence of this anisotropic signal in the
experimental data, and its astrophysical interpretation, is a
fundamental challenge in the study of ultra high energy cosmic
rays (UHECR) and is not free from controversies. The AGASA
experiment reported \cite{hay99} an interesting result claiming a
$4.5\sigma$ excess at energies in the vicinity of $\sim
10^{18}$eV. Although AGASA was located in the Northern Hemisphere
and the Galactic Center was out of its field of view, the
anisotropy signal was located at low galactic latitudes and seemed
to point to the general direction of the Galactic Center. This was
the most significant, but not the only evidence in favor of some
cosmic ray flux excess associated with the Galactic plane. In
fact, Fly´s Eye data \cite{FlysEyeGC} also showed some anisotropy
related to the Galactic plane in a similar energy interval: $4
\times 10^{17} -- 10^{18}$ eV, although not claim was made of a
relationship with the Galactic Center. A little later, the AGASA
anisotropy received further support from a reanalysis of data from
SUGAR \cite{SUGARGC}, which did have the Galactic bulge inside its
field of view, and found a possible point-like excess compatible
with the AGASA signal. However, and rather surprisingly from a
theoretical point of view, the SUGAR point-signal was not located
at the Galactic Center itself, but at $\sim 10^{o}$ from that
position, making any astrophysical interpretation difficult, since
that result excluded the most obvious candidate, the Galactic
supermassive black hole, as a possible source \cite{GMTAAWGC}.
Recently, an analysis using data from the Pierre Auger Observatory
\cite{AGGC05,AGGC06}, which has the Galactic Center well inside
its field of view and has already doubled the AGASA exposure, does
not support the AGASA anisotropy result. It must be noted that
Auger has already collected four times more events than AGASA in
the same energy interval and angular window.

The later Auger result, strengthen by HiRes data that also points
into the same direction (see a HiRes article in this Proceeding),
seems to rule out the AGASA claims. However, it does not rule out
the Galactic Center as a source of EeV cosmic rays. In fact, in
this work we will present an almost unavoidable scenario for the
generation of anisotropy in the cosmic ray signal associated with
the inner regions of our Galaxy. It is based not on a hypothetical
Galactic accelerator, but on well known characteristics of the
extragalactic flux of high energy particles and of the
interstellar medium (ISM) in the Galactic Center instead.
Essentially, the large magnetic fields immersed in the very dense
interstellar medium present in the inner $\sim 200$ pc of the
bulge, partially trap the known extragalactic cosmic ray flux
diffusing into the region and converts into neutrons a sizable
fraction of it. We will also show that the outgoing neutron flux
should be seen as a point source located at the Galactic center.
This neutron source should be observable, at an acceptable level
of statistical significance, during the lifetime of the largest
cosmic ray experiments like, in particular, the Pierre Auger
Observatory.

After a brief review of those characteristics of the Galactic
Center that are specially important for this work, we proceed to
estimate of the neutron production mechanism. Our results are
conclusions are confirmed through analytical, semi-analytical and
detailed Monte-Carlo calculations based on the proposed scenario.
Finally, we also discuss the statistical significance of the
calculated signal for particular case of the Pierre Auger
Observatory.

\section {The Galactic Center} \label{sec:GC}

The center of our Galaxy is a very peculiar region permeated by
magnetic fields in excess of 100 mG that encompasses a very high
density molecular zone. Gas densities in the interstellar medium
can reach up to $\sim 10^{4.5}$cm$^{-3}$ with high filling factor
$>$ 0.1 \cite{ms96}. Through radio maps from the CO
emission\cite{br03}, one can trace the global distribution of
matter inside the region, and create models for the matter
distribution therein. Those observations indicate that the gas
distribution can be approximated by a plane parallel profile
centered around the Galactic plane. The vertical density profile
falls exponentially with a scale height $H_Z \sim 30$pc, from a
maximum value of $\sim 10^{4.5}$cm$^{-3}$ measured at the galactic
plane. This high density, high filling factor region, is
denominated Central Molecular Zone (CMZ). The transition to
conditions more typical of the galactic disk occurs at a
galactocentric distance of $\sim 200$ pc.

The magnetic field structure inside the CMZ can be observed or
inferred using several techniques. The morphology and intensity
can be obtained from infrared polarimetry, Zeeman effect and radio
synchrotron emission data. Those observations demonstrate that
there are at least two distinct regular components, besides the
turbulent one, forming the global magnetic field that permeates
the inner 400 pc of our Galaxy. One of the regular magnetic
components is perpendicular to the galactic plane and is mainly
traced by long filaments of radio emission, denominated
non-Thermal radio filaments. The magnetic field inside those
filaments seems to reach values of the order of some few $\times
10^{3}$ $\mu$G \cite{Novak2003}. Their morphology suggests that
they could be part of a poloidal component of the global magnetic
field, and the fact that one can find those filaments basically
anywhere inside the region can lead to the conclusion that they
could be tracing a large-scale poloidal field, rooted in a central
dipolar momentum, that affects scales comparable to the size of
the whole CMZ. A second component is revealed in polarization data
\cite{Chuss2003}, where magnetic field vectors parallel to the
galactic plane are clearly seen. Those observations suggest that
the Central Molecular Zone is in fact filled by large-scale
toroidal magnetic field, with a global intensity of some few
$\times 10^{2}$ $\mu$G.

Therefore, the field inside the 400pc of our galaxy is between 2
and 3 orders of magnitude larger than in the neighboring Galactic
plane and is remarkably well structured inside the CMZ. The
regular component is a composition of poloidal and toroidal/radial
contributions. Inside Galactic Center there are regions where
either component is dominant, and the combination of those
components strongly suggests a existence of an A0 dynamo operating
inside the CMZ.

\section{Neutron production in the Galactic Center} \label{sec:neutronGC}

The large intensity of the magnetic field (100$\mu$G-3mG) inside a
galactocentric distance of $\sim 200$ pc, increases by a large
factor de crossing time of incoming particles of the high energy
extragalactic cosmic ray flux. Since the density is also very
large inside the CMZ, the trapped particles traverse a much larger
amount of matter than they would otherwise while crossing the
regular interstellar medium that characterizes the rest of the
Galactic plane \cite{km05}.

Thus, the Galactic Center environment could be an efficient
conversion region, where charged protons from the extragalactic
cosmic ray flux could be transformed into neutrons through
hadronic interactions taking place therein. The principal process
should be the reaction $p+p\to n+p+N\pi$ which, besides a neutron
source, would lead to the production of a high energy $\gamma$-ray
source from the $\pi$-decay channels.

In particular the observed magnetic field intensity and spatial
scale could be capable of partially entrapping protons with
energies as high as $10^{19}$eV, but with a greater efficient in
the $<$ 10 EeV energy range. In addition, this is the energy a
neutron should have in order to arrive at the solar circle since,
in the laboratory reference frame, the mean distance it travels
before decaying is $d_{kpc}^{n} \approx 8 \times E_{EeV}$, which
is comparable with the distance between the Sun and the Galactic
center. Thus, mean life time of the neutron helps to tune the
optimum energy to observe the Galactic Center neutron source at
$\lesssim 1$ EeV, since lower energy particles, although being
more numerous in the cosmic ray flux and being more efficiently
trapped in the CMZ, produce very few neutrons that are able to
survive up to the Earth. Neutrons that decay on the fly transform
into protons that are strongly deflected by the Galactic magnetic
field and dilute away in the background cosmic ray flux
\cite{GMTAAWGC}. It is only the surviving neutrons that give a
clear, point-like signal that might be eventually observable with
existing experiments.

\section{Numerical simulations} \label{NumSimulations}

In order to calculate the corresponding neutron flux at Earth, we
developed a Monte-Carlo computer code to propagate particles in
the Galactic Center region with the observed matter and magnetic
field distributions. Those particles were injected randomly at the
boundary of the region, assumed to be a sphere centered in Sgr A,
with a radius of 200pc. The direction of the momenta of the
injected particles at the boundary was also chosen randomly under
the further assumption of an isotropy cosmic ray flux at energies
around 1 EeV.

The intensity of the magnetic turbulence inside the CMZ is not
well established from observations at present. Therefore, in order
to assess the role of the uncertainty in the magnetic field
turbulence on the dynamics of cosmic rays in the ISM of the
confinement region, we perform simulations with three different
field models: (i) purely turbulent, (ii) purely regular and (iii)
an intermediate case, with $\langle B_{turb}\rangle/\langle
B_{reg}\rangle = 1$. In all models, the turbulence was generated
as a superposition of plane waves packets, with a Kolmogorov power
spectrum.

Since there is a considerable amount of mechanical energy being
injected at the center of the region by the several observed
supernova remnants, the approximation used to model the regular
component of the magnetic field inside the CMZ, was inspired on
the Heliospheric magnetic field and it was obtained by the
combination of three contributions: (i) a dipole, (ii) a dynamo
component resultant of a current system that propagates radially
outward along the equatorial plane and closes meridionally over
the spherical boundary of the CMZ and the Galactic polar axis, and
(iii) a radial component resultant from a ring current over the
Galactic plane.

The different realizations of the total magnetic field were
normalized in such a way as to reproduce the average energy
density observed inside the CMZ.

The propagated particles, protons in the resent work, were also
allowed to interact hadronically with the ambient ISM. Hadronic
interactions were treated using the model Sybill 2.1 \cite{en99}.
At this stage, only surviving protons and neutrons were tracked.
After their production, neutrons are propagated through the ISM
out of the CMZ, where they can undergo further interactions with
the ISM. From the external border of the region up to the Earth,
spallation reactions with the ISM are neglected, and only decay is
considered in the model.

\section{Neutron conversion efficiency} \label{sec:Neutron}

We can estimate the neutron flux that is leaving the Galactic
Center from very general arguments. First we estimate the factor
$f$ by which the time that a particle spends inside the
confinement region increases when compared to the free-fly
crossing time. This factor can be written as $f\sim\Delta
t/\tau_{lt}\sim cL/D$, where $\Delta t \approx L^{2}/D$ is the
characteristic diffusion time through a region of size $L$ and
diffusion coefficient $D$, and $\tau_{LT}$ is the light travel
time through that same region.

The total mass traversed by the particle in the region, $\Lambda$,
can be written in terms of the diffusion coefficient:

\begin{equation}
\Lambda \sim f \times m_p \times \langle n_{ISM} \rangle\times
L_{GC}
\end{equation}

\noindent where $\langle n_{ISM}\rangle$ is the average matter
density (e.g., as estimated from the radio emission data) and
$L_{GC}$ is the transversal scale, respectively, of CMZ. Taking
takes, as an order of magnitude estimate, $L_{GC} \sim 200$ pc and
densities varying typically between $10^{4.5}$ and $10^{3}$
cm$^{-3}$ inside the confinement region, one obtains $\Lambda \sim
1-10 \times f$ g/cm$^{2}$ which, considering that the mean free
path of protons in the ISM is $\lambda_{pp} \approx 40$
g/cm$^{2}$, already gives an idea of the relevance of the
phenomenon.

Using the simulation code described in Section
\ref{NumSimulations}, we propagated particles at $10^{18}$ eV in a
purely turbulent magnetic field and in a composite field, in order
to calculate an effective diffusion coefficient ($D_{eff}$) for
cosmic rays crossing the region. From those simulations we
obtained a volume averaged value of $D_{eff}^{turb}\sim 4.6 \times
10^{26}$ m$^2$/s for the purely turbulent scenario and
$D_{eff}^{mix}\sim 6.1 \times 10^{26}$m$^2$/s for the mixed field
scenario.

Therefore, the total mass traversed by a typical cosmic ray proton
at $1$ EeV would be $\Lambda_{SA,turb} \sim 58$ g cm$^{-2}$ for a
pure turbulent magnetic field, and $\Lambda_{SA,mixed} \sim 45$ g
cm$^{-2}$ for a mixed field. Therefore, considering a
proton-proton interaction length $\lambda_{pp} \sim 40$ g
cm$^{-2}$, we see that $\sim 80$\% of the protons entering the
region should experience, on average, one interaction in the
purely turbulent case, and $\sim70$\% in the mixed field. Since
the multiplicity for neutron production in p-p reactions is $\sim
1/4$, the neutron production efficiency per incoming proton can be
estimated as
$\Psi^{[n/p]}\simeq1/4\times\left(1-e^{-\Lambda/\lambda_{pp}}\right)$,
so for both models of the magnetic field, we have
$\Psi^{[n/p]}_{SA,turb}\simeq 0,19$ and $\Psi^{[n/p]}_{SA,mix}
\sim 0,17$ for the pure turbulent and mixed fields respectively.

The previous result can be checked using the Bohm diffusion
approximation,
$D_{Bohm}\sim\frac{1}{3}r_lc\sim3\times10^{25}m^2s^{-1}$).
Furthermore, taking into account that the simulations point to an
effective diffusion length of the order of $0.10 \times L_{GC}$,
the traversed mass is $\Lambda_{Bohm}\sim 90$ g cm$^{-2}$, the
resultant neutron production efficiency is
$\Psi^{[n/p]}_{Bohm}\sim 0,22$.

Table \ref{tab:nprate} shows, on the other hand, the neutron
production efficiency per injected proton into the CMZ.

\begin{table}[htb]
\caption{Neutron production rate for full Monte-Carlo
simulations.} \label{tab:nprate}

\begin{center}

\begin{tabular}{@{}llll}

\hline
Field & $\Psi^{[n/p]}_{MC}$ & $\sigma$\\
\hline  
Reg.&0,38& $2 \times 10^{-3}$ \\
Mixed&0,28& $2 \times 10^{-3}$ \\
Turb.&0,16& $4 \times 10^{-4}$\\
\hline
\end{tabular}
\end{center}
\end{table}

\begin{figure}[htb]
\begin{center}
\includegraphics*[width=8cm,angle=0]{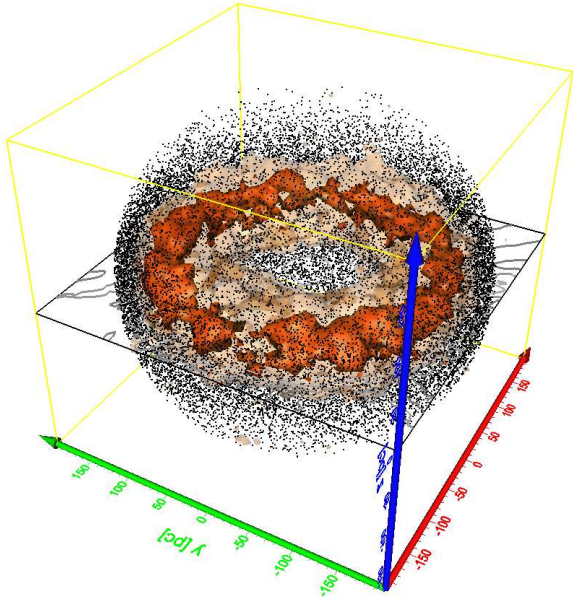}
\caption{Neutron conversion efficiency for the mixed field
scenario. The dots are the production sites for individual
neutrons and two iso-density surfaces are also shown, depicting
the shape and locus of the conversion zone.}
\label{fig:neutron.prod}
\end{center}
\end{figure}

Table \ref{tab:nprate} shows that the purely regular magnetic
field is the most efficient for neutron production. This
characteristic was already expected since, in this kind of field
structure, a large fraction of particle orbits repeatedly crosses
high density regions, inducing neutron conversion. This can be
appreciated in figure \ref{fig:neutron.prod} that, even for the
mixed field scenario, shows that neutron conversion takes place
preferentially at a ring surrounding the Galactic plane near the
external region of the CMZ.

\section{Expected neutron flux at the Pierre Auger Observatory}

From the neutron conversion efficiency in the Galactic Center, one
can calculate the the total neutron flux arriving at the Pierre
Auger Observatory for a given energy. We can assume that the
cosmic ray flux at $10^{18}$eV at Earth is the same that is
crossing the Galactic Center, since at this energy the flux is
believed to be mainly extragalactic. So, the emerging neutron flux
at the Galactic Center can be written as:

{\setlength\arraycolsep{2pt}
\begin{eqnarray}
J_{n,GC}[s^{-1}]\approx 2\pi \Psi^{[n/p]}J_{RC,extgal} \times
A_{GC}
 \label{eqn:jngc}
\end{eqnarray}}

\noindent where $J_{RC,extgal}$ is the extragalactic cosmic ray
flux, $\sim10^{-33.5}cm^{-2}s^{-1}sr^{-1}eV^{-1}$ at $10^{18}eV$,
and $A_{GC}$ CMZ boundary surface area.

So, the neutron flux arriving at a detector at Earth subtending a
solid angle $\Omega$, can be written as:

\begin{equation}
J_{n,det}[s^{-1}]\approx
J_{n,GC}^{[s^{-1}]}\frac{\Omega^{sr}}{4\pi} \label{eqn:jndet}
\end{equation}

However, it is necessary to take into account at least three main
suppression factors. The first is that neutrons are unstable and
some fraction of them decay on the fly between Earth and the
Galactic Center. Taking $t$ as the flying time to earth, and
$\tau$ as their mean life, this factor can be written as:

\[
F_{dec}=e^{-t/(\gamma\tau)} \sim 0,5
\]

The second factor is due to the fact that protons generating
neutrons at $10^{18}$eV must have an energy a larger than
$10^{18}eV$. Therefore, the flux of protons entering the Galactic
Center is smaller than the one previously considered by a factor
that is proportional to the energy. Nonetheless, it is important
to note that the increase in the proton energy is not
significantly high to either alter the simulation dynamics in the
production region or modify considerably the reaction cross
sections.

We can calculate this suppression factor by taking into account
that the cosmic rays flux at this spectral region is a known power
law, $J(E) \propto e^{-2,7}$, and that the average
neutron-to-proton energy ratio in the $pp\to n$ reaction can be
written as:

\[
\frac{\langle E_n\rangle}{\langle E_p\rangle}=\int_0^1 x g(x)
dx=\int_0^1 3x(1-x^2) dx
\]

\noindent where $g(x)$ is the nondimensional inclusive
cross-section \cite{Jones}. So we can write this suppression
factor due to the smaller flux of higher energy protons as:

\[
F_{HEP}=\frac{N(E_n)}{N(E_p)}=\left(\frac{\langle
E_n\rangle}{\langle E_p\rangle}\right)^{2,7} \sim 0,5
\]

The third suppression factor is due to the detection
characteristics of the Pierre Auger Observatory at $10^{18}$eV. As
the observatory is optimized for a higher energy range, at
energies of $\sim 10^{18}$eV, not all impinging cosmic rays are
detected. This factor is:

\[
F_{PAO} \sim 0,5
\]

Therefore, the total suppression flux factor is:

\begin{equation}
F_{Sup} = F_{dec} \times F_{HEP} \times F_{PAO} \sim 0,1
\label{eqn:fsup}
\end{equation}

\noindent which means that only $\sim 10$\% of the neutrons
emitted at the Galactic Center, should be detected at Earth. Thus,
using equations \ref{eqn:jngc}, \ref{eqn:jndet} and
\ref{eqn:fsup}, we estimate in table \ref{tab:fluxaugersup} the
neutron flux detected at the Pierre Auger Observatory in the
energy interval $1-2 \times 10^{18}$eV.

\begin{table}[!htp]

\caption{Expected neutron flux at the Pierre Auger Observatory in
the energy interval $1-2 \times 10^{18}$eV.}

\label{tab:fluxaugersup}

\begin{center}

\begin{tabular}{@{}lc}

\hline
{\bf Field}&{\bf Flux (n/year)}\\
\hline
Regular&39\\
Mixed&29\\
Turbulent&17\\
\hline
Analytical (Bohm)&23\\
Semi-analytical (Mix.)&18\\
Semi-analytical (Turb.)&20\\
\hline
\end{tabular}

\end{center}
\end{table}

An important result in Table \ref{tab:fluxaugersup} is the fact
that, the estimated number of neutrons per year inside the solid
angle subtended by the CMZ is rather independent of the
assumptions made on the field structure and, furthermore, is
comparable with the expectations obtained from simple analytical
and semi-analytical approaches.

\section{Signal statistical significance}

The statistical significance of the neutron signal can be
calculated using the method firstly proposed by Li\&Ma in
\cite{LiMa} for $\gamma$-ray astronomy. Thus,

\begin{equation}
S=\frac{J_{src}}{\hat{\sigma}(J_{src})}=\frac{J_{src}\tau^{1/2}}{\sqrt{J_{sgn}+\alpha^2J_{bg}}}
\end{equation}

\noindent where $J_{src}$ is the flux solely from the source,
$J_{bg}$ is the background flux, $J_{sgn}$ is the total measured
flux (source+background), $\tau$ is the integration time and
$\alpha$ is the fraction of time that the detector observes the
source.

\begin{figure}[htb]
\begin{center}
\includegraphics*[width=8cm,angle=0]{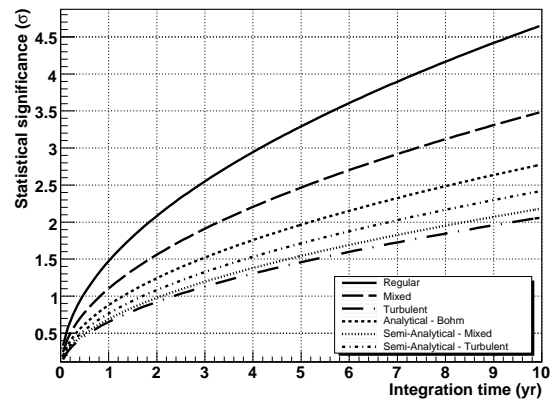}
\caption{Statistical significance of the neutron signal as a
function of integration time for the full Auger detector.}
\label{fig:statsig}
\end{center}
\end{figure}

Therefore, assuming that when the Galactic Center is not in the
sky there is no other point source at those energies, that all the
received signal is background (i.e., the extragalactic cosmic ray
flux) when the neutron source is not on the sky, and taking into
account the previously calculated fluxes for the source at the
Galactic Center, the integration time necessary for the full
Pierre Auger Observatory to detect the neutron source with a given
statistical significance. The results, for all the magnetic field
models considered in this work, are presented in figure
\ref{fig:statsig}, where it can be seen that the the Galactic
Center should be detectable as a point source with at least 2.0
$\sigma$ in 10 years of the complete array operation,
independently of the chosen magnetic field model. Furthermore, in
the more likely case of a mixed field scenario, 5 years should be
enough for a detection at $2.5 \sigma$.

Additionally, a future space experiment like EUSO, with at least
$\sim 10^{5}$ km$^{2}$ of effective area, should be able to detect
this neutron source in less than 2 years at $> 8.5 \sigma$.

\section{Conclusions}

Based on rather simple arguments and very general analytical and
semi-analytical calculations, as well as on detailed Monte Carlo
simulations, it was demonstrated that the Central Molecular Zone
inside the inner 200 pc of our Galaxy should act as a source of
neutrons at EeV energies. Furthermore, this source should be
detectable by the Pierre Auger Observatory within 3-10 years,
independently of the actual magnetic field structure present in
the CMZ, at a statistical significance between 2.0 and 4.5
$\sigma$ in the energy interval $1-2 \times 10^{18}$eV. This is a
lower limit, since only p-p interactions were considered, despite
the fact that p-$\gamma$ interactions could also give a sizable
contribution due to the large infrared background present in the
region. An independent test to the actual existence of this source
would be the detection of TeV gamma radiation consistent with the
p-p interaction rate expected due to partial trapping of cosmic
ray protons inside the CMZ. Detail estimates of this flux
component are under way. This could help explain the $\gamma$-ray
emission detected by the H.E.S.S. collaboration \cite{hess06}.

The Galactic center neutron source could be the first high energy
point cosmic ray source. This neutron source could be useful as a
calibration standard for extensive air shower reconstruction
techniques because of its well defined location on the sky.
Additionally, its detection would enhance our present
understanding of the Galactic Center environment.

\vspace{0.5 cm}

{\it Acknowledgments} This work was partially supported by the
Brazilian Agencies CAPES, CNPq and FAPESP.

\end{document}